# Can Intellectual Processes in the Sciences Also Be Simulated? The Anticipation and Visualization of Possible Future States

Loet Leydesdorff [a]

University of Amsterdam, Amsterdam School of Communication Research (ASCoR), PO Box 15793, 1001 NG Amsterdam, The Netherlands; email: loet@leydesdorff.net

**Abstract**

Socio-cognitive action reproduces and changes both social and cognitive structures. The analytical distinction between these dimensions of structure provides us with richer models of scientific development. In this study, I assume that (*i*) social structures organize expectations into belief structures that can be attributed to individuals and communities; (*ii*) expectations are specified in scholarly literature; and (*iii*) intellectually the sciences (disciplines, specialties) tend to self-organize as systems of rationalized expectations. Whereas social organizations remain localized, academic writings can circulate, and expectations can be stabilized and globalized using symbolically generalized codes of communication. The intellectual restructuring, however, remains latent as a second-order dynamics that can be accessed by participants only reflexively. Yet, the emerging "horizons of meaning" provide feedback to the historically developing organizations by constraining the possible future states as boundary conditions. I propose to model these possible future states using incursive and hyper-incursive equations from the computation of anticipatory systems. Simulations of these equations enable us to visualize the couplings among the historical—i.e., recursive—progression of social structures along trajectories, the evolutionary—i.e., hyper-incursive—development of systems of expectations at the regime level, and the incursive instantiations of expectations in actions, organizations, and texts.

---

[a] I acknowledge Staša Milojević and Andrea Scharnhorst for comments on a previous draft. An initial version of this paper was presented at the workshop "Simulating the Social Process of Science," April 7-11, 2014, at the Lorentz Center in Leiden.

**Introduction**

Agent-based modelling (ABM) has become very popular in the social sciences ever since the publication of Epstein & Axtell's book *Growing Artificial Societies: Social Science from the Bottom Up* (1996). Epstein (2006) formulated his "generativist" research program as a manifesto stating that one cannot explain a social phenomenon until one has "grown" it by simulating the phenomena under study as emerging from the bottom-up. In science & technology studies, this research program accords with the strong program in the sociology of science: individuals and their aggregates in institutions are to be considered as the units of analysis that generate the dynamics of science (Edmonds *et al.*, 2011). From this perspective, the sciences are considered as community-based belief systems, and the units of analysis are agents or collectives driven by a blend of socio-epistemic interests.

Although the agents—scientists—are able to perceive and understand the intellectual dimensions of their sciences, the intellectual organization of the sciences at the supra-individual level is not conceptualized in ABM as the substantive result of discursive interactions. The intellectual organization of the sciences is considered as part of their social organization, and content is defined in terms of the individual cognition of the interacting agents (Payette, 2012; Sun *et al.*, 2013). From this perspective, references and citations can be understood as rhetorical devices in scientific practices (e.g., Cozzens, 1989; Gilbert, 1977; Gilbert & Woolgar, 1974).

ABM has the advantage of being defined in terms of observable behavior; but an "agent-based ontology" (McGlade, 2013, at p. 395) entails problems when simulating mental processes. How can one observe cognition, let alone the group dynamics of negotiations among individual cognitive states (Ahrweiler, 2011)? Krohn *et al.* (1992), for example, simulated knowledge production in research groups as the outcomes of negotiations using the model of laboratory studies (Knorr & Mulkay, 1983).

However, the outcomes of these discussions at the group level eventually have to be written up in manuscripts or working papers containing knowledge claims (Latour & Woolgar, 1979). The drafts have to circulate in other exchange processes at the field level before being sufficiently codified for acceptance as part of the scientific literature (Myers, 1985; Pinch, 1985). These condensates of individual and communicatively shared cognitive resources and processes furthermore contain traces of social organization, and structure the literature in terms of latent factors that can be recognized as fields or disciplines (e.g., Van den Besselaar, 2001). However, the designation of the densities (or other patterns) in the networks that emerge in agent-based simulations always requires a theoretical inference. The interpretation of the patterns is not given naturalistically.

Although focusing equally on the *social* dynamics of science, Sun *et al*. (2013, p. 4) note that "(f)uture 'science of science' studies have to gauge the role of scientific discoveries, technological advances, and other exogenous events in the emergence of new disciplines against the purely social baseline" that resulted from their simulations. However, the authors claim that their account of the emergence of disciplines is the first that can be validated on the basis of empirical data. In a similar vein, Edmonds *et al*. (2011) state that "science is substantially a social phenomenon;" and "agent-based simulations of social processes are able to incorporate lessons from qualitative social science studies of what scientists actually do on a day-to-day level as well as insights from the more naturalistic philosophers of science." Research programs about "the simulation of the social processes of science" have increasingly been organized in terms of workshops,[1] special issues,[2] and edited volumes (e.g., Scharnhorst *et al*., 2012).

In this study, I propose alternatively to consider the sciences not primarily in terms of the belief systems of actors, but as systems of rationalized expectations to which agents and organizations have reflexive access and can thus contribute to their restructuring (Husserl, 1929 and 1935; Luhmann, 1990). Localized manifestations can be considered as instantiations in which puzzles can be solved and new knowledge claims constructed (Giddens 1979). The theory and computation of anticipatory systems enable us to model expectations and then proceed to their simulation (Dubois, 1998; Rosen, 1985).

First, I assume communication as the unit of analysis of the intellectual and social organization of the sciences (Gilbert, 1997). Communications can be attributed to agents as first-order variables; but the results of the interactions among communications (e.g., densities and components) are second-order variables—variables attributed to the first-order variables—that may remain latent for the agents involved, yet structure nonetheless their further communication. In the sciences, the interactions among communications shape discourses that tend to be highly codified, for example, as jargons. Codification is functional for the determination of quality in the context of justification (leading to revision and rewriting) whereas the agents provide the discourses with knowledge claims from below (for example, in observational reports and manuscripts). The structure of communications in science can thus be expected to contain interactions among rationalized expectations, which can only be accessed reflexively by individuals or discussed in organizations.

---

[1] For example, at http://www.lorentzcenter.nl/lc/web/2014/607/info.php3?wsid=607&venue=Oort
[2] For example, at http://simsocsci.blogspot.co.uk/2014/10/cfp-special-issue-of-scietometrics-on.html

At the local level, the complexity of this next-order communication framework can programmatically be reduced to the belief systems of individuals and organizations. This sociological reduction of rationalized expectations to belief systems of communities and agents has been a radical tenet of the so-called Sociology of Scientific Knowledge (SSK) or "the strong program" in science studies (e.g., Barnes, 1977; Barnes & Edge, 1982; Bloor, 1976). However, this perspective over-sociologizes the study of the sciences: it no longer matters whether or not a statement is "true"; but one seeks instead to explain why agents believe that a statement is true in terms of socio-cognitive interests. The codes of the communication are thus no longer attributed in terms of their functionality in the communications, but are operationalized sociologically in terms of interests, for example, of authors, journals, publishers, or other stakeholders. The discourse can then be considered as a rhetorical game driven by career perspectives and institutional interests (Cozzens, 1989; Edge, 1979; Gilbert, 1977).

In my opinion, this focus on belief systems and behavior fails to address the specificity of the modern sciences as cognitive structures of expectations (e.g., paradigms) that are constructed and reconstructed in discursive exchanges in addition to and on top of contingent interests in the context of discovery. The differentiation between rationalized expectations and belief systems was central to the Scientific Revolution of the 17$^{th}$ century (Merton, 1942). Unlike belief systems, which tend to be integrated hierarchically, the sciences are also expected to develop discursively in terms of theoretical and empirical arguments.

The discursive mediation provides a third context to the distinction between the local context of discovery and the global context of justification as formulated in the philosophy of science. Although this latter distinction was reformulated in the sociology of science as the group/field distinction (Rip, 1981; Whitley, 1984), these "dialectical" co-evolution models fail to appreciate a third dynamics of discourse and texts as a context of mediation. This third context of the dynamics in the scientific literature has nonetheless been central to the scientometric enterprise (Price, 1976; Wyatt *et al*., in preparation). Unlike a co-evolution between two contexts, a "triple helix" can endogenously generate crises and other forms of complex dynamics (Krippendorff, 2009; Ivanova & Leydesdorff, 2014; Ulanowicz, 2009).

In response to the strong program, Mulkay *et al.* (1983) first raised the question of "why an analysis of scientific discourse is needed," and Gilbert & Mulkay (1984) then added—on the basis of a laboratory study of oxidative phosphorylation which led, among other things, to Peter Mitchell's Nobel Prize for Chemistry in 1978—that different repertoires co-exist in the sciences. Scholars attribute error to a contingent repertoire and acceptance of knowledge claims to an empirical one, and thus distinguish between truth and error in terms of translations. In the ensuing "sociology of translation," Callon *et al*. (1983 and 1986) suggested considering the sciences as a semiosis—that is, a system of signs in texts (Callon & Latour, 1981; Latour, 1988; Wouters, 1998). Authors, for example, are represented in the text as author names or references, institutional addresses are provided in a byline to the title, and the texts contain words and cited references that indicate the intellectual organization of the arguments.

In a study entitled "In Search of Epistemic Networks," Leydesdorff (1991) argued that three dimensions can be distinguished analytically: authors and their aggregates in communities and institutions, texts which can be aggregated in bodies of literature such as journals and repositories, and cognitions that are organized in theoretical frameworks leading to new ideas, hypotheses, and heuristics. The reduction of cognitive structures to texts or social agency objectifies the sciences and makes them amenable to measurement using geometrical metaphors. However, one should avoid the "reification trap" of an "agent-based ontology" (McGlade, 2013, pp. 294f.) by distinguishing between the operationalization and what is operationalized. Simulations enable us additionally to address the cognitive structures of expectations algorithmically. As Leibniz (1692) put it: "What I love best about the calculus is that (…) it frees us from working with our imagination" (Yoder, 1988, p. 175). The psychological imagination plays with geometrical topologies, but the sciences develop at the supra-individual level and with time as one more degree of freedom.

**Structures of rationalized expectations as anticipatory systems**

Unlike texts and agents or their institutions, cognitions are not given, but remain constructed. These constructs are both organized—for example, in our minds or in their local settings—as instantiations (Giddens, 1979) and self-organizing as "horizons of meaning" (Husserl, 1929). Luhmann (e.g., 1990) added that "truth" can then be considered as a symbolically generalized medium of communication (Parsons, 1968) that enables us to shortcut the communication by using another channel. The symbolic coding of the communication enables us to handle more complexity because the language usage can be restricted (Bernstein, 1971; Coser, 1975).

Within paradigmatic frameworks, for example, one does not need to explicate core assumptions all the time, and thus more complexity can be handled per unit of time. The paradigm or horizon of meaning is globalized, whereas day-to-day activities are localized and thus organized contingently (Fujigaki, 1998). The counter-intuitive point is the possible inversion of the arrow of time, which has deep consequences for the ontology of the cognitive dimension and structures of expectation: coding provides uncertainty with meaning; but meaning is provided to the events from a perspective of hindsight—that is, at a moment $t + 1$ given an event at time $t$. Thus, one looks backward, while the stream of events moves forward.[3]

In their "sociology of expectations," Brown & Michael (2003) formulate this tension as between "retrospecting prospects and prospecting retrospects." In a similar vein, Latour (1987, at p. 97) observed that "the two versions (…) are not uttered by the same face of Janus." In terms of social constructivism, one can also distinguish between the constructing agency and the resulting constructs of prospects, insights, and meanings (Borup *et al*., 2006). The constructs are constructed in terms of codified meanings, and thus have a status different from observable agency or semiosis among texts.

---

[3] One is reminded of Walter Benjamin's ([1940] 1974) "Angel of History."

Meaning is not objective, but subjective and potentially inter-subjective. Codification operates on meaning by reinforcing its backward restructuring of historical events using a dynamics of cultural evolution that touches the historical ground by being instantiated in action. These instantiations can be considered as both retention and historical reproduction mechanisms. The constructs can be shared—taught and learned—using discourse. Husserl (1929) suggested that the intersubjectivity in the constructs is "transcendental" to the contingent events and thus drives the development of the sciences (Husserl, 1938). At the time, Husserl noted the absence of instruments for the specification of these (co-)evolutionary mechanisms:

> We must forgo a more precise investigation of the layer of meaning which provides the human world and culture, as such, with a specific meaning and therewith provides this world with specifically "mental" predicates. (Husserl, 1929, at p. 138; my translation).

Cognitive structures are not objective, but they can be expected to drive the observable instantiations as a virtual order of uncertainties (Giddens, 1979, p. 64; Luhmann, 1990a). What can be considered as feedback and what as feeding forward in these co-evolutions may also change over time. When the system is under (re)construction, agency can be expected to drive the exploration; but the construct can be expected to develop further in terms of its structures, and structures can also be stabilized over time. Whereas structure operates by selecting deterministically from variation (Hodgson & Knudsen, 2011), the structures of expectation contain uncertainty and can thus serve at other moments as sources of variation.

Note that in terms of information theory, a feedback against the arrow of time would generate negative entropy or, in other words, redundancy. However, Shannon (1948) deliberately coupled information theory as probabilistic entropy with thermodynamic entropy—also in order to focus on the electrical-engineering problem and not on the meaning of the communication (p. 3). From this (Shannon) perspective, negative entropy is not possible (Krippendorff, 2009). Following Husserl, however, a Cartesian dualism is introduced between the worlds of uncertainties (information) and meaning: reflexivity (*res cogitans*) operates against the arrow of time, whereas *res extensa* necessarily generates entropy. Meanings can be shared and thus add to the redundancy. The imprint of the *res cogitans* on the *res extensa* can sometimes be measured (Leydesdorff & Ivanova, 2014). But let us remain in this study with the inversion of time as an algorithmic problem.

**Incursion and hyper-incursion in anticipatory systems**

Rosen (1985) defined an anticipatory system as a system that entertains a model of itself. The model provides the system with access to other possible states. Dubois (1998; cf. Dubois & Resconi, 1992) proposed to model the anticipated states using incursive and hyper-incursive equations. In these difference equations, future or present states can be considered counter-intuitively as independent variables driving systems of expectations (Leydesdorff & Franse, 2009).

One can reformulate recursive equations into (hyper-)incursive ones by changing the temporal parameters: instead of $x_t$ as a function of $x_{t-1}$, one writes $x_t$ as a function of $x_{t+1}$. Dubois' prime example is the logistic equation that can be used, among other things, for modelling biological phenomena. This so-called Pearl-Verhulst equation is formulated as follows:

$$x_t = ax_{t-1}(1-x_{t-1}) \tag{1}$$

An incursive version of this equation, *mutatis mutandis*, reads as follows:[4]

$$x_t = ax_{t-1}(1-x_t) \tag{2}$$

and the corresponding hyper-incursive model is formulated by Dubois (1998) as:

$$x_t = ax_{t+1}(1-x_{t+1}) \tag{3}$$

Whereas the logistic equation models a growth process with a feedback in the case of biology, the incursive equation models an instantiation at the present moment as a process that is hyper-incursively embedded in the structure of expectations provided by Eq. 3.

In the case of Eq. 2, for example, the anticipatory system $x$ builds on its previous state ($x_{t-1}$), but it selects among its current options and thus realizes one instantiation among other possible states. From this perspective, the use of the recursive formulation of the equation (Eq. 1) for modeling social phenomena can be questioned as a biological metaphor. The market as a social system of expectations, for example, does not select commodities, technologies, etc., from among options provided at a previous moment (that is, $1 - x_{t-1}$), but selects from options in the present while restructuring itself as an instantiation on the basis of previous states.

---

[4] Another incursive equation is . This quadratic equation has two roots [ and $x = 0$], which correspond to the steady states of Eq. 2 to be discussed below (Leydesdorff & Franse, 2009, at pp. 110f.).

In Eq. 3, history ($x_{t-1}$) no longer plays a role, but the system is overwritten at $x_t$ in terms of future states. Note that future states are only available as expectations, and only a model of the system enables us to use expectations for the restructuring of the system. Thus, these equations address core notions of how the sciences operate at the epistemic level. How can predictions entertained in models drive the sciences?

Dubois (2002, pp. 112 ff.) makes a further relevant distinction between weakly and strongly anticipatory systems. A weakly anticipatory system entertains a model to predict future states of a system under study; a strongly anticipatory system uses future states for restructuring itself in a process of self-organization. Whereas individuals can entertain a model of the system reflexively as weakly anticipatory systems, the systems of rationalized expectations—the horizons of meaning—are themselves restructured in terms of refinements of the expectations and can thus be considered as strongly anticipatory and as operating at an intersubjective level.

**Solving the equations**

The incursive and hyper-incursive equations have other solutions than the recursive ones. Let us first solve the incursive Eq. 2:

$$x_{t+1} = ax_t(1 - x_{t+1}) \qquad (2)$$

$$x_{t+1} = ax_t - ax_t x_{t+1} \qquad (2a)$$

$$x_{t+1}(1 + ax_t) = ax_t \qquad (2b)$$

$$x_{t+1} = ax_t /(1+ax_t) \qquad (2c)$$

By replacing $x_{t+1}$ with $x_t$ in Eq. 2c, two steady states can be found for $x = 0$ and $x = (1 - a)/a$, respectively. These steady states correspond to the non-existence of the system ($x = 0$) and a line in the bifurcation diagram of $x$ against the parameter $a$. In Figure 1, this line of the steady state is penciled on top of the well-known bifurcation diagram of the recursive formulation of the logistic equation (e.g., May, 1976). As is well-known, the bifurcation diagram of the recursive (Pearl-Verhulst) equation is increasingly chaotic when $a \to 4$, and cannot exist for $a \geq 4$. However, the incursive system can be instantiated both in the domain of $a < 4$, and in the non-biological (e.g., psychological) domain of $a \geq 4$.

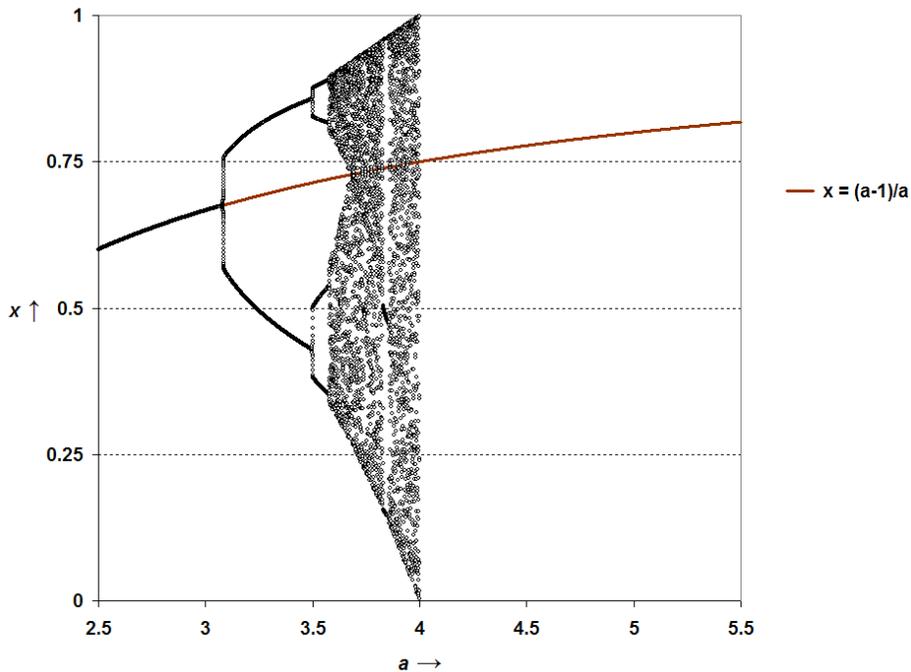

**Figure 1**: The steady state of the weakly anticipatory system. (Source: Leydesdorff & Franse, 2009, p. 111).

The line penciled into Figure 1 can perhaps be considered as an emerging axis stabilizing an *identity* among the reflections at each moment of time (Leydesdorff & Franse, 2009). The weakly anticipatory system provides meaning to the events by integrating them in both the biological domain ($a < 4$; e.g., bodily perceptions) and the social domain of meaning-sharing and processing ($a \geq 4$). The instantiation of the different representations integrates them historically and thus functions as a historical linchpin (*cogito*) for developing a strongly anticipatory system in the cultural (i.e., non-natural) domain of meaning-processing ($a \geq 4$). However, the latter domain can be considered as a *cogitatum*, that is, part of a reality about which one can be expected to remain in doubt (Luhmann, 1990a).

Equation 3 is a quadratic equation with two solutions:

$$x_t = ax_{t+1}(1 - x_{t+1}) \tag{3}$$

$$x_t = ax_{t+1} - ax_{t+1}^2 \tag{3a}$$

$$ax_{t+1}^2 - ax_{t+1} + x_t = 0 \tag{3b}$$

$$x_{t+1}^2 - x_{t+1} + x_t / a = 0 \tag{3c}$$

$$x_{t+1} = \tfrac{1}{2} \pm \tfrac{1}{2} \sqrt{[1 - (4/a)\, x_t]} \tag{3d}$$

This system has no real roots for $a < 4$, but it has two solutions for values of $a > 4$. (For $a = 4$, $x = \tfrac{1}{2}$.) These solutions are added to Figure 2.

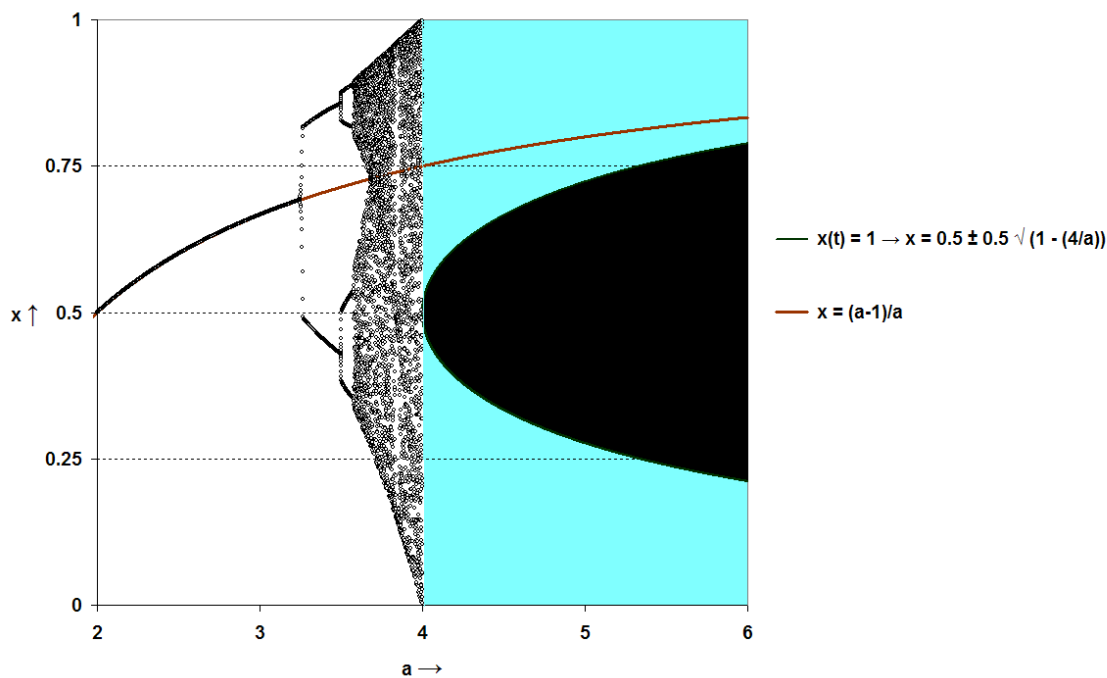

**Figure 2**: The system of expectations as a result of hyper-incursion (Leydesdorff & Franse, 2009, at p. 113).

For $a \geq 4$, two expectations are generated at each time step depending on the plus or the minus sign in the equation. After $N$ time steps, $2^N$ future states could be possible if this system would operate without historical constraints. Thus, the system of expectations continuously needs a mechanism for making decisions between options because otherwise this system would rapidly become overburdened with uncertainty. In other words, the communication cannot further be developed without a form of agency making choices between options because of the continuous proliferation of uncertainty by the hyper-incursive mechanism.

Decisions by agency can anchor the anticipatory system historically in instantiations. Luhmann (2000) suggested considering decisions as the structuring mechanism of organizations. Although this reflexive capacity is conceptualized by Luhmann as endogenous to the communication system, organizations can also be attributed to social structure as institutional agency. From this perspective, a psychological system can perhaps be considered as the minimal unit of reflection for making choices (Habermas, 1981; Leydesdorff, 2000). Both agency and organizations are able to integrate perspectives by reflexively making choices.

If decisions are socially further organized—for example, by using decision rules—an institutional layer can increasingly be shaped. The institutional layer provides a retention mechanism for the next round of expectations (Aoki, 2001). Thus, the system can be considered as dually layered, as (*i*) a forward-moving retention mechanism and (*ii*) sets of possible expectations which flow through the networks. Note that the possible expectations can be expected to proliferate much faster than the retention. Unlike the instantiations, these "horizons of meaning" are not given, but continuously in flux and undergoing constant reconstruction (Luhmann, 2002). While the agents and the texts are both part of the recursive retention mechanism (*res extensa*), the agents can additionally be expected to act incursively (as the *cogitantes* in *res cogitans*).[5]

**Simulations**

The equations remain very abstract because the referents of *x* are not yet specified. The advantage of this abstractness is that this referent can also be cognitive as against textual (e.g., co-words) or social (e.g., co-authors). For example, the latent dimensions of networks or the development of eigenvector centrality can also be modeled (Leydesdorff, 2010). But how can one move from these abstract bifurcation diagrams to modeling the epistemic dimension of the sciences operating as strongly anticipatory systems?

---

[5] In the semiotic tradition—actor-network theory and the sociology of translation—a distinction is made between agents in sociology and "actants" in the narrative (e.g., Latour, 1996).

Using simulations, Leydesdorff (2005) showed in a first step that an incursive routine can generate an observer endogenously. Challenged by a competition for visualizations organized by Katy Börner at the time (2007; http://vw.indiana.edu/07netsci/), a colleague suggested exploiting the graphical interfaces of Windows to input a recognizable picture into the simulations of incursive and hyper-incursive routines so that one would be able to recognize observationally what these routines do to the representations. Let us postpone the discussion of whether and how knowledge can be represented as a dynamic, and focus first on what these routines can do when using a stabilized representation.

In a computer language—I will use Visual Basic 6 below—one can consider a picture or any representation as an ordered set of pixels that can each be transformed in terms of their respective colors using the computer code provided in Table 1 as an example for the recursive case (to be discussed below). *Mutatis mutandis*, the simulation can be extended for incursive and hyper-incursive equations, and in a next step one would even be able to specify interaction terms between pixels in several routines disturbing one another. But before we complicate the issue further, let us explore an example.

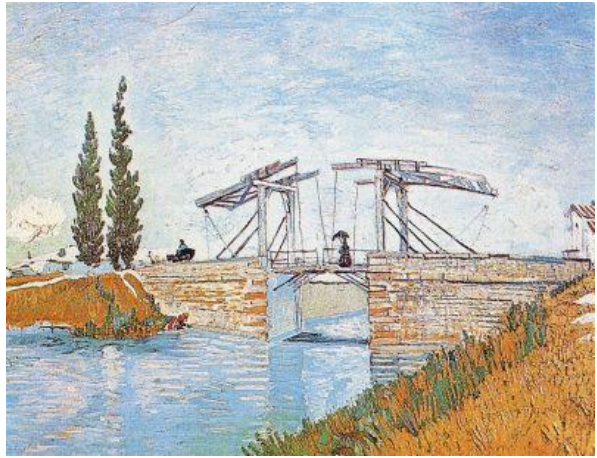

**Figure 3**: Van Gogh's "Langlois Bridge at Arles" to be used as input to the routines.[6]

Figure 3 shows Van Gogh's well-known "Langlois Bridge at Arles" that I will use as an exemplary representation in the routines. The height of this reproduction is 308 pixels (3322 twips) and the width is 400 pixels (4200 twips). Visual Basic uses twips because this measure is screen-independent. In the simulation of the bridge at Arles, I will use an array of 3322 * 4200 twips, or approximately $14 * 10^6$ data points.

Using the recursive formulation of the logistic equation (Eq. 1), for example, one can expect a transformation of this representation for $1 < a < 3$, an oscillation for $3 \leq a < 3.57$, and further bifurcation and development towards chaos for larger values of *a* between 3.57 and 4. As argued above, this will be very different for the incursive and hyper-incursive formulations of this same model.

---

[6] This image is in the public domain; see at http://commons.wikimedia.org/wiki/File:Vincent_Van_Gogh_0014.jpg .

**Table 1**: Transformation of a representation using the logistic equation (Eq. 1).

```
1   DO
2       For Y = 0 To PicFrom(0).ScaleHeight
3           For X = 0 To PicFrom(0).ScaleWidth
4               ' Get the source pixel's color components.
5               clr = PicFrom(0).Point(X, Y)
6               r = clr And &HFF
7               g = (clr \ &H100) And &HFF
8               b = (clr \ &H10000) And &HFF
9
10              'scale between zero and one
11              rt = r / 256
12              gt = g / 256
13              bt = b / 256
14
15              'transform incursion
16
17              rt = (param * rt) * (1 - rt)
18              gt = (param * gt) * (1 - gt)
19              bt = (param * bt) * (1 - bt)
20              r = Int(rt * 256)
21              g = Int(gt * 256)
22              b = Int(bt * 256)
23
24              If r > &HFF Then r = &HFF
25              If g > &HFF Then g = &HFF
26              If b > &HFF Then b = &HFF
27
28              ' Write the new pixel.
29              clr = RGB(r, g, b)
30              PicTo(0).PSet (X, Y), clr
31          Next X
32          DoEvents
33      Next Y
34      ' Make the changes permanent.
35      ' PicTo(0).Picture = PicTo(0).Image
36      PicFrom(0).Picture = PicTo(0).Image
37      cmdGo.Enabled = False
38      MousePointer = vbDefault
39
40      DoEvents
41      If param2 = 0 Then Exit Do
42
43  Loop While True
```

Table 1 provides the code for the core transformation of the colors in the recursive case. Two pictures are first distinguished: PicFrom(0) with horizontal (*x*) and vertical (*y*) values in lines 2 and 3, and PicTo(0) in lines 35 and 36 for the resulting picture. One can recognize the logistic equation in lines 17-19 for the red, green, and blue components of the twips at each specific position. I use the traditional RGB (red-green-blue) decomposition for the colors. Since the logistic equation (Eq. 1) requires values for *x* between zero and one, the color values are first divided by 256 in lines 11-13, and then renormalized for the picturing in lines 20-22.

The bifurcation parameter *a* is provided interactively by the user, and is called "param" in lines 17-19. The DoEvents in lines 40-41 makes the program sensitive to switching to another routine or quitting. The program then runs in two loops for the horizontal *x* (line 31) and the vertical *y* (line 33), respectively. Each time it runs, the original picture (PicFrom) is replaced by the newly generated one. For example, the representation can be expected to erode in a number of steps towards chaos for values of *a* > 3.57 when using the logistic equation recursively (Eq. 1). Figure 4, which can be run interactively using the program available at http://www.leydesdorff.net/simulation.2015/netsci.exe , combines the recursive, incursive, and hyper-incursive routines in a single setting.

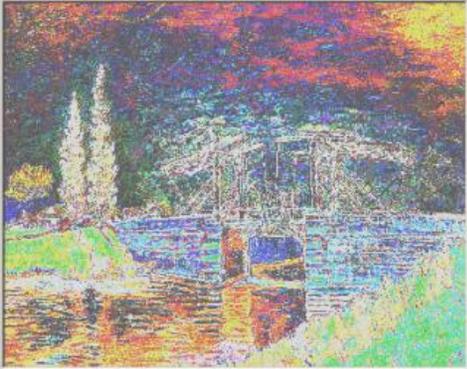

**Figure 4**: Recursion, incursion, and hyper-incursion in cases of using the logistic equation with Van Gogh's "Langlois Bridge at Arles;"
http://www.leydesdorff.net/simulation.2015/netsci.exe
21

Figure 4 shows the different states of the system after a number of runs with the bifurcation parameter $a = 3.6$, as specified by the user in the box at the top right. Since $a > 3.57$, the representation is decaying using the logistic equation in the left-top (PicFrom) and middle-top (PicTo) representations that alternate (since after each loop PicTo becomes defined as PicFrom).

I added two reflexive observers using the incursive routine. They are first able to hold their representation by building on the previous state, but secondly by observing in the present. The first observer is generated in the left-bottom screen observing directly the original picture (PicFrom) in the left-top screen. (For a more extensive discussion of the generation of an observer and observers observing each other see also Leydesdorff (2005) and Von Foerster (1982).) The observer generated in the right-bottom screen, however, does not observe the original picture, but only its transformation using the hyper-incursive equation as depicted in the screen box in the middle at the bottom. Note that the representation in this latter box seems almost to have disappeared, but the observer is able to regenerate it. We thus show the possibility of transmission of the observation at a distance. My argument is that this is not a social process, but a communication process. The state of mind of the observers and their social contexts are not relevant to the reception, which is determined by the communication of representations.

**Discussion and further perspectives**

These simulations are far from perfect; they can be considered as conceptual ("toy") simulations that enable us to explore further options (McGlade, 2004, p. 118). However, the reproduction of the Van Gogh's bridge at Arles provides a primitive representation when compared with the complexity conveyed in knowledge representations. I could have used a more complex representation such as a chemical structure, but even then, the knowledge is not in the picture, but in the argument. A static representation is by definition an instantiation (McGlade, 2003). However, one can read the argument in a scientific text sometimes in terms of a sequence of figures and tables, and visualizations and animations can be used to summarize an argument. How to represent the knowledge itself has hitherto remained a largely unsolved problem. Perhaps one could also work with (e.g., Bayesian) probability distributions in terms of grey-shades, but the representation would then probably remain more abstract (Leydesdorff, 1992).

My argument has been one against the reification of the cognitive process, but not against operationalization. By distinguishing what is operationalized from what is operationalizing, one gains reflexivity about the representation. Different reference worlds can be indicated by using different equations. Whereas the recursive equation models the development of a system (e.g., a biological population) in *res extensa*, the observer at the left-bottom of Figure 4 can be considered as an individual observer using an incursive routine. The hyper-incursive routine, however, indicates an evolution that is no longer rooted historically, but can only be entertained reflexively by the second observer since it refers to an intersubjective communication domain of expectations. As Luhmann (1984, at p. 226; 1995, 164) formulated: "The most important consequence of this analysis is *that communication cannot be directly observed, only inferred. To be observed or to observe itself, a communication system must be flagged as an action system.*" Luhmann added a reference to Warriner (1970, p. 106), who formulated: "The basic problem in the theory of communication lies in the general reluctance of the social scientist to deal with what is not directly observable."[8]

The knowledge dimension can sometimes be measured as redundancy or, in other words, the relative absence of uncertainty after specification of the *negative* selection mechanisms that can be expected to operate on the observable variation. Leydesdorff & Bensman (2006), for example, showed that the intellectual organization of citation patterns among journals is not to be retrieved in the power-law distribution of the majority of the citation relations, but in the well-known "hook" of this curve representing the most frequent relations—that is, in the area where the Poisson process is interrupted for intellectual reasons (Milojević, 2010). Negative entropy or redundancy is not observed directly, but can be measured as a negative imprint (Leydesdorff & Ivanova, 2014). Retrieving this imprint requires theoretical specification, which generates the cognitive domains of expectations that are here under study.

The simulations in this study are technically a first step and mainly meant to demonstrate a possible direction for further research. In the above simulations, for example, one could only change a single bifurcation parameter interactively. The recursive, incursive, and hyper-incursive routines were run with the same bifurcation parameter, while I argued also that only the incursive equation has a solution for all values of $a$. When one feeds the current routine, for example, with $a = 6$, the recursive representation immediately becomes chaotic; but if one first uses the "Go" at the hyper-incursive routine in the middle at the bottom, one obtains an oscillation. It would be worthwhile to explore the further option of using different bifurcation parameters for the various routines.

---

[8] In his later work, Luhmann (e.g., 1999) formulated a theory of observation different from Maturana (e.g., 1977) and following Spencer Brown (1964) and Von Foerster (1993); cf. Leydesdorff, 2006.

Another further extension would be to add the possibility that the observers interact in a network. Using models for adoption, one could, for example, add the possibility that observers accept a majority vision given an environment as in the case of a "lock-in" (Arthur, 1989; Leydesdorff, 2001). For example, one could add to the above simulations a cellular automaton generating variation and structures at the same time. How do the two layers of intellectual and social organization relate, and under what conditions can control shift from local action to global communication, and *vice versa*? As we know, networks are constructed bottom-up; but since the coding in the networks develops as eigenvectors (Von Foerster, 1993), these latent dimensions can be expected to feedback and structure the room for contributions. Because the latent *cogitata* remain uncertain and not given, the structuring cannot become deterministic, but remains "structuration": one is not controlled, but "constrained and enabled" by the emerging codes of communication (Giddens, 1979).

In other words: while the intellectual, textual, and social organization of the sciences can be expected to co-evolve, these co-evolutions are far from symmetrical. The codes are attributes of the communications, which are in turn attributes of the communicators. Thus, the intellectual networks can develop in terms of a second-order dynamics. Intellectual expansion can therefore develop much faster than the social conditions (Weinstein & Platt, 1973). As Slezak (1989) noted in a debate with proponents of the strong program, one should not even try to explain the larger variance of cognitions in terms of the smaller variance in social organizations.

The sociological design can also be inverted: scholars often do not generate much content in communication, but intellectual content is the reason for communication. The new content enters the discourse as a proposal for a rewrite (Fujigaki, 1998). The structure is selective, and thus the social processes tend to follow the intellectual ones. As I argued in the introduction, this presumes the symbolic generalization of codes of communication and their differentiation into a modern and pluralist society (Luhmann, 1990b and 2002a; Merton, 1942). In older types of society (so-called "high cultures"), functional differentiation among the codes of communication tends to be suppressed and hierarchical belief systems tend to prevail. Self-organization is based on an additional degree of freedom for restructuring the historical organization of meaning reflexively.

The use of incursions and hyper-incursions in scientific communication inverts the axis of time and thus the entropy flow can be negative. Leydesdorff & Ivanova (2014) have elaborated on the mutual redundancy in three or more dimensions generated by the possibility to share meanings in the communication of information. The same information can be provided with different meanings; in other words, the sharing generates redundancy. Using codes of communication, one can additionally translate by relating different meanings and thus generate a third level in the communication: first communication is *relational*; second, systems of communication *position* the information and thus enable us to provide meaning to information; and thirdly, meanings can be *translated* (Callon, 1986). One thus obtains a much richer picture of scientific communication that includes the possibility to explore the operation and interaction of codes of communication in relatively shielded niches (Petersen *et al*., in preparation).

In terms of the philosophy of science, I have followed Luhmann's (1986) intuition that Husserl's transcendental concept of "intersubjectivity" can be operationalized in terms of communication (Knudsen, 2006; Paul, 2001; cf. Schütz, 1952). Although Luhmann (e.g., 1990, p. 113, n. 59) indicated that codes of communication can be considered as functionally differentiated using the metaphor of eigenvectors which stand orthogonal,[9] the notion of anticipation and expectation could not be operationalized further at that time (Leydesdorff, 2012, p. 88).[10] However, the theory and computation of anticipatory systems provide us with the tools to take this next step toward the simulation of the sciences as structures of expectations.

---

[9] Luhmann (1990, p. 113, n. 59) indicates the latent dimensions with the word "eigenvalue" formulating as follows: 'I remain with the terminology of "eigenvalues," although one should distinguish "eigenvalues" from "eigenstructures" and "eigenbehavior".' Technically, the eigenvalue of an eigenvector is the factor by which the eigenvector is scaled when multiplied by the matrix.

[10] A reference to Rosen (1985) can be found in Luhmann (1997, Vol. 1, at p. 206n.), but the concept of "anticipation" is used in this context only metaphorically.

Leydesdorff, L., & Bensman, S. J. (2006). Classification and Powerlaws: The logarithmic transformation. *Journal of the American Society for Information Science and Technology, 57*(11), 1470-1486.
Leydesdorff, L., & Franse, S. (2009). The Communication of Meaning in Social Systems. *Systems Research and Behavioral Science, 26*(1), 109-117.
Leydesdorff, L., & Ivanova, I. A. (2014). Mutual Redundancies in Inter-human Communication Systems: Steps Towards a Calculus of Processing Meaning. *Journal of the Association for Information Science and Technology, 65*(2), 386-399.
Luhmann, N. (1984). *Soziale Systeme. Grundriß einer allgemeinen Theorie*. Frankfurt a. M.: Suhrkamp.
Luhmann, N. (1986). Intersubjektivität oder Kommunikation: Unterschiedliche Ausgangspunkte soziologischer Theoriebildung. *Archivio di Filosofia, 54*(1-3), 41-60.
Luhmann, N. (1990a). The Cognitive Program of Constructivism and a Reality that Remains Unknown. In W. Krohn, G. Küppers & H. Nowotny (Eds.), *Selforganization. Portrait of a Scientific Revolution* (pp. 64-85). Dordrecht: Reidel.
Luhmann, N. (1990b). *Die Wissenschaft der Gesellschaft*. Frankfurt a.M.: Suhrkamp.
Luhmann, N. (1995). *Social Systems*. Stanford, CA: Stanford University Press.
Luhmann, N. (1997). *Die Gesellschaft der Gesellschaft*. Frankfurt a.M.: Surhkamp.
Luhmann, N. (1999). The Paradox of Form. In D. Baecker (Ed.), *Problems of Form* (pp. 15-26). Stanford, CA: Stanford University Press.
Luhmann, N. (2000). *Organisation und Entscheidung*. Opladen: Westdeutscher Verlag.
Luhmann, N. (2002a). The Modernity of Science. In W. Rasch (Ed.), *Theories of Distinction: Redescribing the descriptions of modernity* (pp. 61-75). Stanford, CA: Stanford University Press.
Luhmann, N. (2002b). The Modern Sciences and Phenomenology. In W. Rasch (Ed.), *Theories of Distinction: Redescribing the descriptions of modernity* (pp. 33-60). Stanford, CA: Stanford University Press.
Maturana, H. R. (1978). Biology of language: the epistemology of reality. In G. A. Miller & E. Lenneberg (Eds.), *Psychology and Biology of Language and Thought. Essays in Honor of Eric Lenneberg* (pp. 27-63). New York: Academic Press.
May, R. M. (1976). Simple mathematical models with very complicated dynamics. *Nature, 261*(June 10), 459-467.
McGlade, J. (2003). The map is not the territory: complexity, complication, and representation. In R. A. Bentley & H. D. G. Maschner (Eds.), *Complex Systems and Archaeology* (pp. 111-119). Salt Lake City, UT: University of Utah Press.
McGlade, J. (2014). Simulation as narrative: contingency, dialogics, and the modeling conundrum. *Journal of Archaeological Method and Theory, 21*(2), 288-305.
Merton, R. K. (1942). Science and Technology in a Democratic Order. *Journal of Legal and Political Sociology, 1*, 115-126.
Milojević, S. (2010). Modes of Collaboration in Modern Science: Beyond Power Laws and Preferential Attachment. *Journal of the American Society for Information Science and Technology, 67*(7), 1410-1423.
Mulkay, M., Potter, J., & Yearley, S. (1983). Why an Analysis of Scientific Discourse is Needed. In K. D. Knorr & M. J. Mulkay (Eds.), *Science Observed: Perspectives on the Social Study of Science* (pp. 171-204.). London: Sage.
30